\documentclass[a4paper]{jpconf}
\usepackage{graphicx}
\usepackage[export]{adjustbox}
\usepackage{float} 
\usepackage{xurl} 
\usepackage[margin=1in]{geometry}
\usepackage{hyperref}
\hypersetup{
    colorlinks=true,
    citecolor=blue,
    linkcolor=blue,
    filecolor=blue,      
    urlcolor=blue,
    }

\begin{document}

\title{Broadening the scope of Education, Career and Open Science in HEP}

\author{Sudhir Malik$^1$, David DeMuth$^2$, Sijbrand de Jong$^3$, Randal Ruchti$^4$, Savannah Thais$^5$, Guillermo Fidalgo$^1$, Ken Heller$^6$, Mathew Muether$^7$, Minerba Betancourt$^8$, Meenakshi Narain$^9$
Tiffany R. Lewis$^{10,11}$, Kyle Cranmer$^{1,2}$ and Gordon Watts$^{1,3}$}

\address{$^1$Physics Department, University of Puerto Rico Mayaguez, Mayaguez, PR 00682, USA}
\address{$^2$Science Department, Valley City State University, Valley City, ND 58072}
\address{$^3$Physics Department, Radboud University, Nijmegen, The Netherlands}
\address{$^4$Physics Department, University of Notre Dame, IN 46556, USA}
\address{$^5$Physics Department, Princeton University, Princeton, NJ 08544, USA}
\address{$^6$School of Physics and Astronomy, Minneapolis, MN 55455, USA}
\address{$^7$Department of Mathematics, Statistics and Physics, Wichita State University, Wichita, KS 67260, USA}
\address{$^8$Fermilab, PO Box 500, Batavia, IL 60510, USA}
\address{$^9$Department of Physics, Brown University, Providence, RI 02912, USA}
\address{$^{10}$NASA Postdoctoral Program Fellow}
\address{$^{11}$Astroparticle Physics Laboratory, NASA Goddard Space Flight Center, Greenbelt, MD 20771}
\address{$^{12}$Department of Physics, New York University, New York, NY 10003, USA}
\address{$^{13}$Department of Physics, University of Washington, Seattle, WA 98195, USA}

\ead{sudhir.malik@upr.edu}

\renewcommand{\footskip}{10pt}  

\begin{abstract}
High Energy Particle Physics (HEP) faces challenges over the coming decades with a need to attract young people to the field and STEM careers, as well as a need to recognize, promote and sustain those in the field who are making important contributions to the research effort across the many specialties needed to deliver the science.  On one hand, trends in education and career opportunities in HEP have been highly selective and generally exclusionary of participation by underrepresented minority and female students. Missed are a significant pool of students and faculty at Predominantly Undergraduate Institutions (PUIs) and Community Colleges (CC) that could be tapped to increase and strengthen underrepresented participation. On the other hand, a current mindset in the field is that highly specialized skills such as software and firmware development and instrumentation development are not broadly recognized as “physics” work, and thus the perception of what it means to ‘be a physicist’ must also be challenged.  The view of what constitutes significant contribution to the physics must be reexamined, expanded and rewarded.  Confronting and addressing these issues would encourage an influx of new workforce into the field, help retain those who are in the field, and equip those who might seek careers outside of HEP. Despite the challenges, there are reasons for optimism. The collaborative nature of the HEP community and the excitement of the science can serve as potentially powerful organizing mechanisms to confront systemic issues such as these. The PUI and CC faculty could be invited to participate and contribute meaningfully to HEP along with their students, who would acquire a variety of skills that prepare them for STEM careers. Access to a master’s degree in Applied Physics could enable technological careers in laboratories, accelerator facilities, the private sector and could open avenues to more advanced degrees including PhDs.   The Open Science tools established by the HEP community can greatly enable such participation. A major effort in software training has been organized and underway through IRES-HEP, to address the breadth and critical nature of the analytical challenges facing HEP over the next two decades.  Inside the HEP community, Grid Computing, Multi-threaded Software, Machine Learning in physics analysis and the development of new detector technologies have empowered our ability to quickly process petabytes of data, see patterns in them, and potentially discover new physics.  And such technological and analytical benefits are generally extendable to other fields including medicine.  Computing, software, detector design, logical thinking, statistical analyses, simulation, and hardware development skills have made physicists highly employable within and outside academia. Such skills can also serve as attractors for students who may not want to pursue a PhD in HEP but use them as a springboard to other STEM careers.  And importantly, an increased pool of interested and engaged students and faculty can spawn new and as yet unforeseen  opportunities for the field.  This paper  reviews  the  challenges  and  develops  strategies  to  correct  the disparities to help transform the particle physics field into a stronger and more diverse ecosystem of talent and expertise, with the expectation of long-lasting scientific and societal benefits.
\end{abstract}

\section{Introduction}
Broader participation by diverse demographic and geographic groups is key to building strength in STEM fields in general, and High Energy Physics (HEP) is no exception. A recent report from the National Science Board Vision 2030~\cite{NSF2020_vision} urges that faster progress in increasing diversity is needed to reduce the significant talent gap in the United States (U.S.). It also indicates that foreign-born individuals are significant contributors to U.S. Science and Engineering. The data shows that a global science and engineering enterprise is growing faster than before, while the U.S. share of scientific discovery is dropping. The silver lining is that the U.S. has a significant lead in fundamental research, and we must innovate and embrace ways to be more inclusive and diverse to maintain our global leadership. For U.S. HEP to stay competitive and evolve stronger, we must address these issues as part of the planning process for the next decade so that an influx of new workforce in the field can be encouraged, retaining those who are in the field and as well equip them with a skills to find careers both within and outside HEP.

The current trend of education, diversity, gender and career opportunities in HEP~\cite{ICHEP_div} is highly selective, and generally exclusionary of participation by underrepresented minority and female students. To address this issue, meaningful opportunities for broader participation are required at different levels. While a proposed Master’s Degree in Applied Physics could enable a technological career in laboratories, accelerator facilities, private sector, or possibly a PhD degree, it can also attract students from less favored communities who may wish to transition to a career with a Masters degree. 

At Ph.D. and post-doctoral level, a vast majority of researchers who choose to stay in academia are hired as teaching faculty at  Predominantly Undergraduate Institutions  (PUIs)~\cite{pui, pui_list} and Community Colleges (CCs). Moreover, these institutions host a vast majority of students from underrepresented groups. The faculty of these institutions, despite their talents, are at disadvantage when it comes to continuity in research that requires commitment of both time and resources. Typically, higher teaching loads, lack of informed guidance,and lack of collaborative opportunities diminish their chances for being competitive for research funding. 

 The strength of the particle physics community lies in its openness, diversity, and access to some of the best creative talent worldwide. This strength has resulted in the advancement of human knowledge that is leading to a detailed understanding of the inner workings of the universe. Engaging undergraduate institutions and their faculty and students in this front-line research enterprise would both enrich and strengthen the workforce and the field. 

With the advent of tools for Open Science, we are poised for an even greater and more significant role, and in a more inclusive way. For example, CERN Open Data~\cite{cern_opendata} is being harnessed to connect discovery science with the public through outreach activities, techniques such as Machine Learning, and much more. 
Development of state-of-the-art particle sensors often find applications outside of particle physics, e.g. in medical imaging, material diagnostics, etc. Small scale development and application research is well in reach of smaller research groups with limited resources, if they can be teamed up with larger scale HEP collaborations.
The software training in the HEP community is transforming the preparation of our next generation of problem solvers by reaching students beyond HEP.

We must strengthen existing successful methodologies, many of which have been created and developed by the U.S. HEP community, and explore new ways of transferring HEP resources, talent, and data into an Open Science paradigm. By doing so we propel our field forward, and garner the interest and support of general public through the next phase of scientific discoveries. 
Simultaneously, we must prepare HEP Ph.D. students  and postdocs for careers inside or outside academia. 

We recognize that advancements and discoveries in physics have become a huge enterprise necessitating advances in computing, software, AI, Machine Learning and detector technologies in order to develop, design, build and operate experiments.   And these consume a significant fraction of time of physicists. This means that our perception of what it means to ‘be a physicist’ and what work is considered a contribution to the physics must evolve and expand. In the following sections we discuss several proposals to address and mitigate the aforementioned challenges.

\section{Assessing the needs of Particle Physics}

The Standard Model of particle physics provides an an impressive description of much of observed particle physics. Nevertheless, it has significant shortcomings and cannot be the ultimate theory of nature at the fundamental scale. Theoretical particle physics provides a plethora of extensions and modifications of the Standard Model, many of which predict new physics phenomena for which there are either minimal or no observational motivations at present. Further groundbreaking experimental programs will be required to make further scientific discoveries.  These can be forged by going to higher energy, measuring more precisely or by performing measurements with new techniques.  It is not unreasonable to assume that new discoveries will incorporate all three of these attributes.  

In all cases the progress will be technologically driven, in accelerator technology, in sensor technology, or in data treatment and analysis technology. 
Strong injection of technological know-how is therefore a requirement for progress. Much of the technological; progress does not necessarily require extensive particle physics knowledge. Rather it does require a deep understanding of technology and the ability to innovate and create new technology.

Unlocking the technological work-force needed for future particle physics research requires recruiting talent outside the normal particle physics oriented study and career paths.
There is potentially large pool of students and faculties at PUIs and CCs that can be tapped to increase such participation. The faculty could contribute to HEP along with students who would acquire various skills that prepares them for STEM careers.

That this potential is not well accessed at the moment is due to resource limitations in these faculties, in terms of research time and equipment and travel budgets. There is also little awareness of the potential for significant contribution from these faculties with the faculties themselves and with the current main players in HEP, the R1~\cite{pui} universities and research institutions.

In sections below, suggestions are made to implement programs to unlock the potential for technical education and research connected to experimental particle physics.  While such a programs would contribute to the development of new workforce to meet the technical challenges of experimental particle physics, they would also contribute to making the HEP community more diverse and offer new opportunities for participation from those underrepresented in STEM. The next two decades of the Snowmass2021 vision can afford an opportunity to review the challenges and develop strategies to correct the disparities and transform the particle physics field into a stronger and more diverse ecosystem of talent, expertise and public-minded support. Working together across the various groups of the Community Engagement Frontier and in consort with the scientific, technical and computation frontiers of the Snowmass2021 Community Self-study, the community can develop an understanding and project the person-power needs of the field in terms of students, physicists, and technical, analytical, and managerial eperts on an annual basis – and in parallel – project the societal benefits that could be acquired through this process. 

Among the facets of this proposed study:
\begin{itemize}
 \item Needs Assessment of the Community: Work with the various research and technical
frontiers to develop projections of their person-power and educational needs on an
annual basis over the next two decades.
 \item Needs Assessment of the Education Process: Review of teaching at all levels –
particularly transition points between education levels – where “transmission
coefficients” of students are low and develop ideas to improve these circumstances.
Understanding these are particularly important to increasing the participation of
underrepresented groups in the field.
 \item Model Development to Guide Community Planning: Engage with AIP and APS who
follow the education process of younger students upward through schools to develop
statistical models of what currently is the situation and how it might be improved, not
only to the benefit of particle physics, but also the broader STEM community.
\end{itemize}

\section{Faculty Collaboration across Academia}

The field of Particle Physics has successfully brought many young researchers through the post-doctoral ranks, at which time these individuals are then searching for more permanent positions in academia, National Laboratories, and in the private sector. In the domain of academia, many individuals may consider opportunities at Carnegie classified R1 universities, other institutions including undergraduate-serving colleges and universities, as well as community colleges. 

For those seeking what are typically teaching-rich positions, the potential exists for these persons  to attract a broader geographic and demographic base of students to the particle physics field which affords the possibility of strengthening the participation of underrepresented groups in the particle physics research program. Engaging those teaching at undergraduate institutions in front-line particle physics research will strengthen our academic workforce, and in turn inspire their students.

Recommendations:
\begin{itemize}
\item Conduct an analysis to assess the needs of faculty at undergraduate institutions in order for them to be successful in developing vibrant research programs, whilst maintaining a significant teaching load characteristic of such institutions.
\item Survey institutional collaborations of PUI and CC faculty with R1 and laboratory groups that have proven successful so far, to assess lessons learned. 
\item Conduct a study of new models of collaboration or cooperation that would allow PUI/CC faculty and their students to collaborate in demonstrably effective ways in experiments, and for faculty to be effective leader rather than regarded as secondary.
\item Conduct a survey of R1 institutions and research laboratory physicists who might share an interest in collaborating with PUI and CC faculty.
\item Identify and inventory past and present activities by PUI and CC faculty engaged in HEP or related activities such as  
experiment design, construction, installation, and maintenance.
\end{itemize}

\section{New Masters Degree in Applied Physics} 

 A Master’s Degree in Applied Physics is proposed here with the aim of providing an advanced degree beyond bachelors level that could open meaningful career paths for students into technological careers in laboratories or the private sector; additionally such a degree could serve as a springboard for those who find that a PhD degree is possible.  This would provide a meaningful opportunity for participation by underrepresented groups and enrich and broaden participation in STEM careers.

Recommendations:
\begin{itemize}
\item Conduct a needs analysis to assess the value and chances of success of such a program.
\item Compare such a program with current models of Masters Degrees in engineering and MBA degrees.
\item Conduct an assessment of private sector buy in and support for such a degree opportunity for current students and for their employees seeking career improvement.
\item Assess the effectiveness of such a degree to  provide training opportunities for those seeking technical careers at laboratories in our field or elsewhere or perhaps medical fields.
\item Identify the curriculum for such degree(s) and whether they could be made available nationwide through shared curriculum.
\item Assess university, PUI and CC buy in and participation in such a program.
\end{itemize}

\section{Training the HEP community}

\begin{figure}[h]
\begin{center}
\includegraphics[width=35pc]{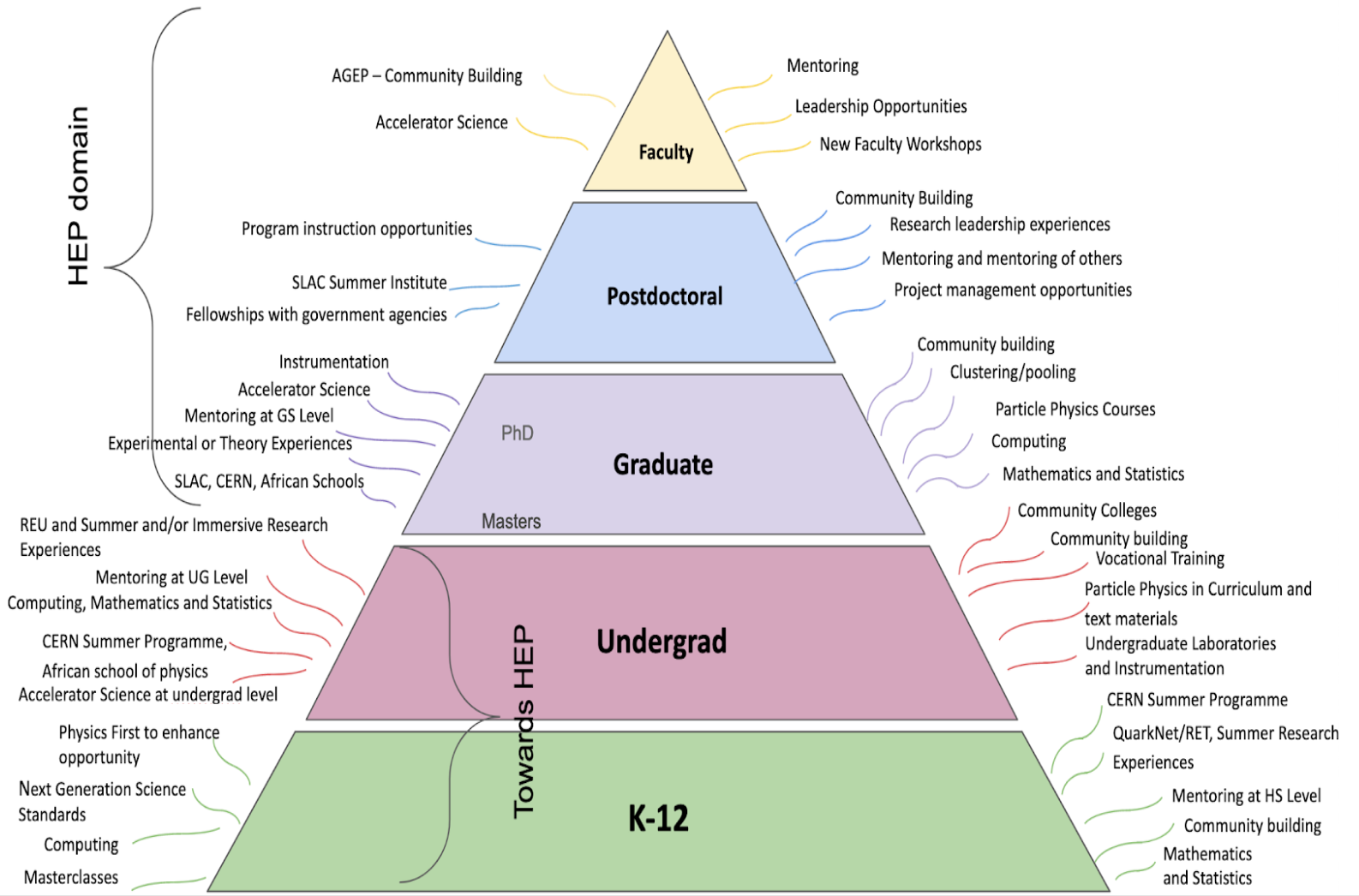}\hspace{5pc}%
\begin{minipage}[b]{28pc}\caption{\label{fig:pyramid}Evolution of HEP Education and Training}
\end{minipage}
\end{center}
\end{figure}

Software, computing, hardware and analysis skills are an integral part of the toolkit of successful HEP experimentalists, and maximizing the science from the investments in current and future HEP projects relies critically on it. This is also a skill set that is readily transferable in case of non-HEP career evolution of people trained in HEP. For example, software training is now key in many research fields, but most users learn software skills only after joining a research program. Individual universities do not uniformly provide such training to students prior to their beginning their Ph.D. research. HEP Computing and hardware skills at university level are even more rare given significant costs and time investments. In addition, domain-specific aspects add challenges to the learning process. Embarking on a HEP-specific training could still involve experiment-specific  environment challenges. Figure~\ref{fig:pyramid} shows the HEP related education and research training evolution from a student to faculty level and with examples of  training and  programs  indicated at different levels. It is important to point out that we must find ways to make sure that diversity, equity and inclusiveness percolates all the way upwards from the bottom to the pyramid of Figure~\ref{fig:pyramid}. 

Some of the current training programs in the field of HEP are described below. It would be very beneficial if some of such training could accorded course credit, even if taken during the summer, so that not only highly competitive students but a bigger subset of students is encouraged and able to take part. And the scientists, faculty, and postdocs involved engaged in these training's should be given compensation for their significant time investment. 

\subsection{Software Training}
No one size fits all while imparting software training. A possible HEP wide solution is to exploit our large-scale community structure and organize training within our research domains~\cite{training_springer}. Recent successful and ongoing efforts like IRIS-HEP~\cite{iris_hep}, FIRST-HEP~\cite{first_hep} and HEP Software Foundation (HSF)~\cite{hsf_hep}  have taken strong and effective steps in this direction and impart training to those within the field and ancillary fields. The HEP community must play a role in bringing together a focus on these efforts to engender sustainability and scalability.  It is important to initiate learning early on in the process of preparing a software-equipped future particle physicist.

At PUIs and CCs, and often at R1 universities, students pursuing undergraduate degrees in physics do not necessarily get exposed to all the core undergraduate physics and computing courses that are related with research. When going on to go graduate school, these students may lag behind their peers. While the curricular issue of what core physics courses are taught at student's institution may be challenging to address, nevertheless we can and should develop a way for students to take elective Particle Physics courses that give them university credit, and in addition provide HEP training in related software. While students may exit at different levels in the HEP education and training pyramid of Figure~\ref{fig:pyramid}, we should try to make sure that opportunities for training reach all.

The software training programs of IRIS-HEP and HSF have trained over 1500 undergrads, Masters, PhDs, postdocs and senior physicists with software skills over the last 3 years, even in times of COVID-19 challenges. One of the commendable achievements of software training has been the building of community around LHC experiments~\cite{training_lhcexp} and DUNE~\cite{training_dune}. Nuclear Physics, which has similar educational needs, is also engaged with their effort led by Jefferson Lab (TJNAF). 

\subsection{Training Internships} 

Internship programs by DOE~\cite{needs_doeintern} and NSF/REU~\cite{needs_nsfreu} have been very popular to encourage students to gain valuable research experience and are highly competitive.  For example DOE sponsored internships like SULI~\cite{needs_suli} place undergraduate physics or engineering majors in paid  summer internships. These internships~\cite{needs_fnalintern} offer a chance for students and professionals (school teachers and visiting faculty) to work with scientists or engineers on projects at the frontier of scientific research in particle physics. 
For all students, but particularly those from academic departments that are primarily PUI or masters degree oriented and which often attract community college transfers, the opportunities afforded by extended visits to national laboratories are truly life changing.  Efforts by Wichita State University to attract and train students for STEM careers is an excellent example of such efforts on behalf students by dedicated and creative faculty. 

\subsection{Experiment Training Efforts}

Several experiments have their own unique programs to train new collaborators  - PhDs, postdocs or faculty in all aspects of experiment. The CMS Experiment~\cite{training_cms,training_cmsdas} and its contingent of US physicists (USCMS) that maintains the Fermilab LHC Physics Center (LPC)~\cite{training_lpcfnal} and an independent LHCb Experiment program~\cite{training_lhcb} are examples of very successful stories. The LPC has recently started a Summer Undergraduate Research Internship program to train students on HEP specific topics like detector hardware, physics analysis and software. Through such a program, students interested in HEP can be trained early on and the skills are translatable to other STEM areas, should the students choose to go in other career directions. 

It is notable that the R\&D and construction phase of a typical HEP experimental project can span several years and often a decade or more, and affords a fertile opportunity to learn detector hardware related to HEP.  And, while major work happens at the laboratory or hosting site of the experiment, substantial building and testing of detector components can be distributed across the collaborating institutions,  This can be broadly beneficial to faculty and students across a wide range of participating colleges, universities and laboratories. 

\begin{figure}[h]
\begin{center}
\includegraphics[width=\textwidth, frame]{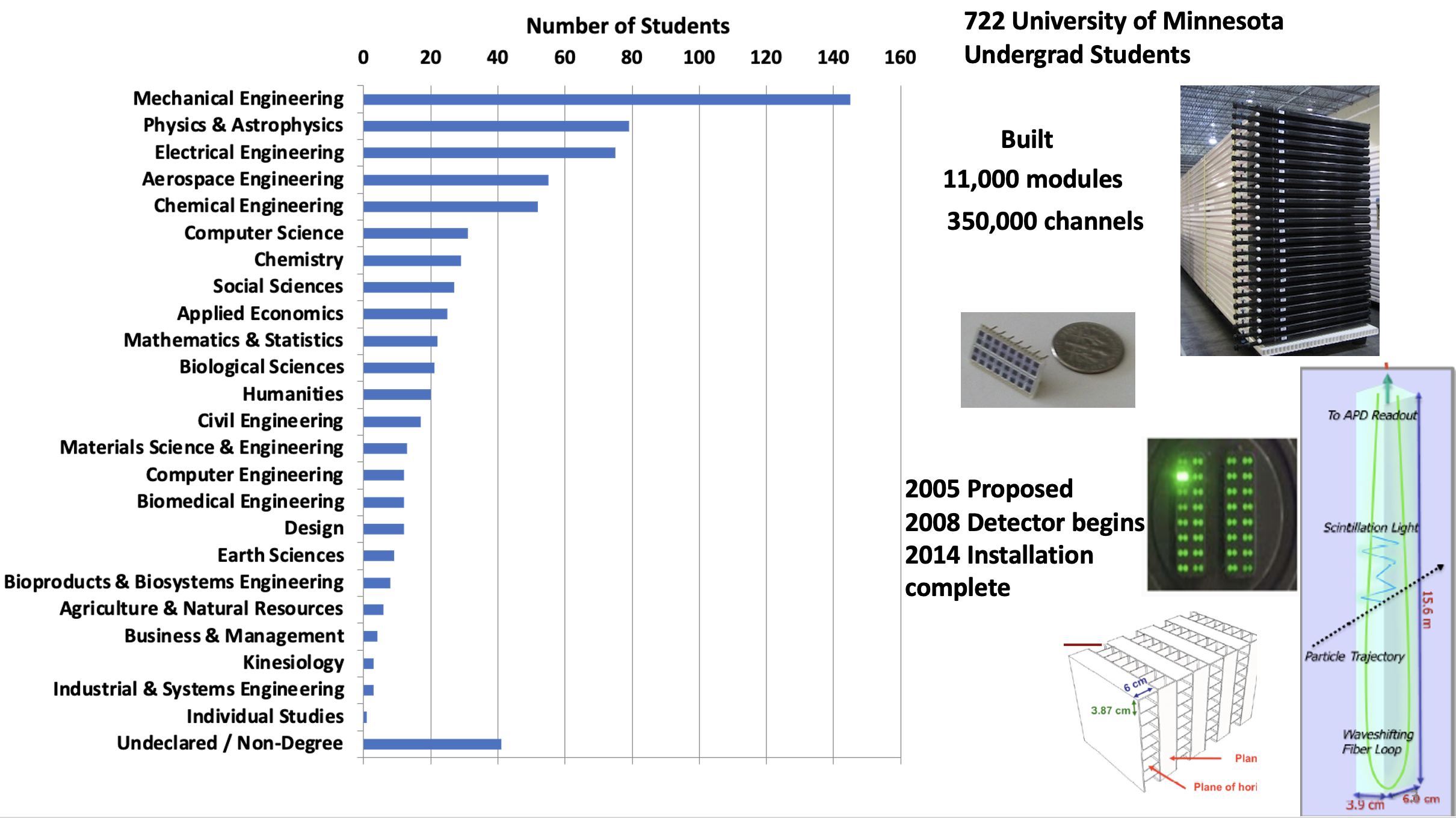}\hspace{5pc}%
\begin{minipage}[b]{28pc}\caption{\label{fig:nova}Over 700 undergraduate students from different disciplines built NO$v$A Detector at University of Minnesota}
\end{minipage}
\end{center}
\end{figure}

Ongoing construction of the CMS Phase-2 Pixel sub-detector~\cite{training_cornellintern,training_cornellpixel} for the CMS experiment at Cornell University and sub-detectors for MINOS~\cite{training_minos}, NOvA~\cite{training_nova} and Mu2e~\cite{training_mu2e} experiments at Minnesota are some of the very good examples that involve hundreds of students over the course of construction from different STEM backgrounds at the host and collaborating  institutions. Figure~\ref{fig:nova} shows the spectrum of undergraduates and their major fields of study, who were involved in the building of the NOvA detector~\cite{training_novavideo}. That experience demonstrated that undergrads can be actively and effectively engaged, alert to and fully capable of recognizing anomalies and solving problems. The Students and postdocs appreciated being supported  as valued team members in the detector building tasks, and for the hands on experience of how to work collaboratively and manage the work of the various teams. Graduate students and postdocs learned about the challenges of building a detector, managing others, and paying attention to time and budget. Moreover, this type of experience is compelling to a general audience and newsworthy, as one reporter captured, “I worked in a coffee shop--this is quite different. I wanted to work in an environment where I could live and breathe science and to be around people who’ve had similar experiences and classes.”

\subsection{Undergraduate Research at PUI's} Initiating meaningful opportunities for research early in a student's career can often lead to the life and career-altering realization that a student actually has the capacity to be a scientist. That all-important self-identification can be the result of well coordinated programs with committed faculty mentors, who are teamed with graduate students and postdocs.  And access to suitable facilities is also a signature of an effective program.  Features of many research invested PUI's are the commitment of a cadre of mentors to attract students to the excitement of the science and attendant research opportunities, and notably, and the participation of engaged students, the most accomplished among them as good as any University's best. As access to research grade facilities may be more challenging for PUIs, we support proposals to develop PUI-centric undergraduate research collaborations, especially in EPSCoR states~\cite{EPSCoR_DOE}~\cite{EPSCoR_NSF}, where program leadership and coordination of HEP related research projects evolve to capitalize on the diversity of talent at the nation's smaller and likely more rural universities and colleges.

\section{Open Science to build HEP community}

The process of sharing scientific findings in journals has been the mode of Open Science~\cite{os_forum} for many decades~\cite{os_broadaccess}, and  Particle Physics and Astronomy were among the first fields to embrace this approach, fostering and sharing open source software, open access to published work and open data for researchers and citizen scientists. The World Wide Web, Invenio, Indico, Inspire, Zenodo and  OSG (Open Science Grid) are examples of some of the platforms and tools provided by the domain of particle physics to the scientific community and beyond. CERN~\cite{os_access} provides access to open source hardware, open publishing and open data via the CERN Open Hardware License, the Sponsoring Consortium for Open Access Publishing in Particle Physics (SCOAP3~\cite{os_physics})and the Open Data Portal for the LHC experiments. The strength of the particle physics HEP community lies in its openness, diversity and access to some of the best creative talent worldwide. With the advent of tools to operate Open Science, we are poised for an even greater and more significant role, and in a more inclusive way. 

\subsection{Benefits}

Open Science empowers the ability to collaborate, question and contribute to sustainable scientific knowledge and process, and accelerate future discoveries. Open data from LHC~\cite{cern_opendata} is employed by educators, students, and self-learners for teaching, learning, and research. It allows research to maximize its impact by reaching a wider audience of participation and sharing the findings, seeking collaborations without borders, and expanding funding and career advancement opportunities. It is also greatly enhances the quality and opportunities for education by making it more affordable and equitable.

\subsection{Challenges}

With opportunity comes responsibility. The particle physics community has become fully aware, and needs to make the outside world equally aware, that there are practical limits to making all raw data public.  Besides problems of facilitating training the public at large to analyze data, significant additional resources are needed to support personnel to maintain the data and to serve as consultants as outsiders attempt to use the associated software tools for analysis, which are not typically static over experimental lifetimes, and evolve as experimentation proceeds.  Existing computing infrastructure would not be able to support many others to access all of the available raw data.

Recommendation:

\begin{itemize}
\item The HEP community should define, with cogent arguments, what should be the scope of making our data and resources publicly available, and the hardware, software and person-power costs associated with such implementation.
\end{itemize}

\subsection{Goals}

We must strengthen existing successful methodologies, many of which have been created and developed by the U.S. HEP Community, and explore new ways of translating HEP resources, talent and data into Open Science.
We can advance this powerful framework to create more inclusiveness by Open Software Training and Open Physics teaching that will propel our field and the general public into the next phase of discoveries.  

\section{Intersection of Physics and Machine Learning}

Machine Learning (ML) is becoming an integral part of physics research. Many critical HEP algorithms for triggering, reconstruction, and analysis rely on ML and there are entire conferences and summer schools dedicated to this crosscutting field. Moreover, the unique constraints of physics experiments and the ability to exploit symmetries inherent in physics data have made the field of physics-informed ML a vibrant sub-field of Computer Science (CS) research. However, despite the relevance and importance of this research, pursuing a career at the intersection of these fields remains tenuous and undefined endeavor. We must  work towards finding ways about how we can better support academic and career development at the intersection of physics and ML, and ensure that we can continue to benefit from and contribute to state-of-the-art ML.

\subsection{Current Environment} 

HEP research has long required advanced, cutting-edge computing techniques, and physicists have historically contributed to the development of these methods. Over the past decades, there have been great advances in the data-processing power of ML algorithms and these methods have been quickly adopted by physicists to address the unique timing, memory, and latency constraints of HEP experiments~\cite{ml_latency}.

Physics is also impacting ML research. The constraints of HEP experiments and known symmetries of physical systems create a rich environment for the development of novel ML (see for example~\cite{ml_neural}). There are even entire conferences and workshops dedicated to this intersection including the Microsoft Physics ML lecture series~\cite{ml_microsoft} and the ML and the Physical Sciences workshop at NeurIPS~\cite{ml_workshop}.

However, despite the demonstrated criticality and vibrancy of this research, from a physics perspective the path to a sustainable research career at the intersection of physics and ML is unclear at best. In fact, many researchers seem to regard ML research and applications as minor side projects\footnote{Due to the lack of community statistics on this topic, the evidence presented here is primarily anecdotal}.

There are very few physics department supported courses on ML\footnote{although~\cite{ml_forphy} provides a good example of how such a course might work} and interested students are often forced to convince CS departments to allow them to attend graduate courses (which typically do not discuss physics-informed ML). Furthermore, at many universities the lack of collaborative interactions between Physics and CS departments limits opportunities for physicists to engage with cutting-edge ML research and develop partnerships with industry researchers\footnote{NYU~\cite{ml_nyu} and UW~\cite{ml_washington} Physics collaborations with the Center for Data Science is an excellent example of cross-department collaboration}. 

As a physics graduate student at a US institute it is essentially impossible to focus one’s thesis research on ML, as most departments do not acknowledge this as ‘physics work’ (despite the fact that these algorithms are in many cases critical to the functioning of physics experiments and data processing). Perhaps most egregiously, although there is now some support for this work at the post-doc or research staff level (see IRIS-HEP~\cite{iris_hep} and IAIFI~\cite{ml_iaifi}), early-career researchers are often actively advised against applying for such positions as they constitute ‘career death sentences’.

\subsection{Proposals for Future Work} 

It is clear that these issues must be addressed in order to best support and advance our field and our researchers. An important first step is to work towards a cultural shift around the perception of what it means to ‘be a physicist’ and what work is considered and accepted as valid as contribution to physics. We must value technical and software contributions in the same manner we value analysis work and foster a supportive and educational environment for graduate students and early-career researchers interested in ML and physics. It benefits no one to discourage individuals from pursuing work in this vibrant and promising field.

Concretely, this necessitates developing cross-department courses, workshops,and research collaborations. Graduate programs should approach technical and computing skills with the same rigor they do traditional physics courses. Departments and advisors should be encouraged to allow students and post-docs to work across disciplines and explore opportunities for industrial partnerships or internships. This process should also include a reevaluation of requirements for physics graduate theses.

The research landscapes of both physics and ML are shifting, and by adjusting to these changes we can ensure a vibrant future for data-intensive physics and physics-informed ML.

\section{Conclusions}

This paper addresses current needs in education and training in High Energy Physics and how the field can be proactive and innovative in transforming current efforts, which tend to be highly selective and generally exclusionary in nature, into a more broadly participatory and inclusive approach, to the benefit of all students and particularly those from underrepresented demographic groups.  There are significant pools of students and faculty across academia and notably those at Predominantly Undergraduate Institutions and Community Colleges that can be tapped for participation, to the benefit of a stronger and more vibrant HEP field for the future. We can develop this untapped talent by providing creative and innovative opportunities in HEP specific education at undergraduate and Masters levels, and by developing rewarding research opportunities through improved training in physics, instrumentation R\&D and software and analytical techniques. Such efforts will not only ensure that the HEP field has the professional workforce needed to meet the scientific challenges and discoveries that lie ahead, but also the capability to prepare students effectively for productive careers in other STEM fields and across the needs of our broader society.

\section*{References}

\typeout{}
\bibliography{iopart-num}


\end{document}